# Effect of Strontium substitution on the structural and magnetic properties of La$_{1.8}$Sr$_{0.2}$MMnO$_6$ (M = Ni, Co) layered manganites

Syeda Karimunnesa[a, b], Bashir Ahmmad[c] and M. A. Basith[a*]

[a]Department of Physics, Bangladesh University of Engineering and Technology, Dhaka, Bangladesh
[b]Department of Physics, Chittagong University of Engineering and Technology, Chittagong, Bangladesh
[c]Graduate School of Science and Engineering, Yamagata University, 4-3-16 Jonan, Yonezawa 992-8510, Japan

## Abstract

Sr substituted perovskites La$_{1.8}$Sr$_{0.2}$MMnO$_6$ (M = Ni, Co) were synthesized using solid state reaction technique to present a systematic study on their morphological, structural and magnetic properties. The average grain size of the as-prepared La$_{1.8}$Sr$_{0.2}$NiMnO$_6$ samples are in the range of 0.2~0.7 μm and those for La$_{1.8}$Sr$_{0.2}$CoMnO$_6$ manganites are 0.1~2.8 μm, which is significantly less than that of unsubstituted La$_2$NiMnO$_6$ (LNMO) and La$_2$CoMnO$_6$ (LCMO) manganites. The XPS analysis enlightened about phase purity, binding energy and oxygen vacancy of La$_{1.8}$Sr$_{0.2}$MMnO$_6$ manganites. The Sr substituted LNMO has revealed a sharp ferromagnetic to paramagnetic phase transition at 160±2 K which is about 120 K less than that of parent LNMO. The Sr substituted LCMO exhibited such a transition at 220±2 K which is 8 K less than that of parent LCMO. The temperature dependent magnetization measurements suggest that the effect of Sr on the transition temperature in LNMO is more significant than that of LCMO.

## Introduction

Perovskite manganites exhibit a wide range of functional properties, such as colossal magneto-resistance, magnetocaloric effect, multiferroic property and some interesting physical phenomena including spin, charge and orbital ordering. The double perovskites La$_2$MMnO$_6$ (M = Ni, Co) are reported to exhibit ferromagnetic behavior and a magnetodielectric effect near room temperature [1-6], making them a promising material for the next-generation memory devices. A relatively high ferromagnetic Curie temperature (T$_C$) is found to be strongly dependent on the M$^{2+}$ and Mn$^{4+}$ cations ordering, as well as on the M-O-Mn superexchange interaction.[4, 7] The perovskite manganites classes of materials La$_2$NiMnO$_6$ (LNMO) and La$_2$CoMnO$_6$ (LCMO) have attracted considerable interest in recent years due to their rich physics [8-12] and potential applications [13, 14] in spintronic devices. [15-17] The initial studies on perovskite manganites system were mainly focused on LaMnO$_3$ which is an antiferromagnetic insulator. [18] It was reported that substitution of La by divalent cations like Sr, Ca and Ba in LaMnO$_3$ [18, 19] introduces mixed valence on Mn ions and initiate the conventional double-exchange (DE) mechanism. [20, 21] Doping the insulating LaMnO$_3$ material, in which only Mn$^{3+}$ exists, with the divalent ions (Ca, Ba, Sr, etc.) causes the conversion of a proportional number of Mn$^{3+}$ to Mn$^{4+}$. The presence of Mn$^{4+}$, due to the doping, enables the itinerant (e$_g$) electron of a Mn$^{3+}$ ion to hop to the neighboring Mn$^{4+}$ ion via DE,

Contact: M. A. Basith ✉ mabasith@phy.buet.ac.bd

which mediates ferromagnetism and conduction. However the substitution of La by other divalent cations in perovskite manganites LNMO and LCMO are comparatively less explored. Therefore, in the present investigation, we intend to explore the effect of Sr substitution in LNMO and LCMO manganites systems. Here we have prepared and studied the effects of Sr addition on the structural, morphological and magnetic properties of $La_{1.8}Sr_{0.2}MMnO_6$ manganites synthesized by conventional solid state reaction technique.

**Experiment details**

The polycrystalline samples having compositions $La_{1.8}Sr_{0.2}MMnO_6$ (M = Ni, Co) were synthesized by using standard solid state reaction technique, as described in detail in our previous investigation. [22] High purity oxides of $La_2O_3$ (99.9%), $SrCO_3$ (99.9%), NiO (99.9%), $Co_3O_4$ (99.9%) and $MnCO_3$ (99.9%) powders of Sigma-Aldrich, UK, were used as raw materials. The reagents were carefully weighed in proper stoichiometric proportion and mixed thoroughly with acetone. Then it was grounded in an agate mortar for 6 h until a homogeneous mixture was formed. The compacted mixtures of reagents taken in desired cation ratios were calcined at 1100 °C for 12 h in a programmable furnace. The calcined powders were grounded for 6 h by an agate mortar and pestle to get a homogeneous mixture. After the calcinations, 10 mm diameter and 1 mm thick pellets were prepared under a pressure of 12000 P.S.I and finally the samples were sintered at 1300 and 1400 °C for 6 h in air with heating rate 3°C/min for investigations.

The crystal structure of the samples was determined from X-ray diffraction (XRD) data. The XRD patterns were collected at room temperature using a diffractometer (Rigaku Ultimate VII) with $CuK_\alpha$ ($\lambda$ = 1.5418 Å) radiation. The microstructure of the surface of the pellets was observed by a field emission scanning electron microscope (FESEM, JEOL, JSM 5800) equipped with the energy dispersive X-ray spectroscopy (EDX). The EDX analysis has been used to determine the overall chemical homogeneity and composition of the samples. X-ray photoelectron spectroscopy (XPS, ULVAC-PHI Inc., 1600) analysis was carried out with a Mg-$K_\alpha$ radiation source. The temperature and field dependent magnetization measurements were carried out both at zero field cooling (ZFC) and field cooling (FC) processes [23] using a Superconducting Quantum Interference Device (SQUID) Magnetometer (Quantum Design MPMS-XL7, USA).

**Results and discussions**

The XRD patterns of $La_{1.8}Sr_{0.2}MMnO_6$ (M = Ni, Co) are presented in Figure 1. It is observed that all the samples show good crystallization with well-defined diffraction line. All the peaks are indexed in the figure and these peaks are typical for the single phase rhombohedral structure. It was also observed that the positions of the peaks comply with the reported values in $La_{1.8}Sr_{0.2}NiMnO_6$ [24, 25] and $La_{1.8}Sr_{0.2}CoMnO_6$ manganites. [26] Earlier, the unsubstituted polycrystalline $La_2MMnO_6$ (M = Ni, Co) sample has revealed a monoclinic structure. [26, 27-28] The lattice parameters calculated from the patterns are displayed in Table 1. From the Table 1, it is seen that lattice parameters of 'a' and 'b' in $La_2MMnO_6$ (M = Ni, Co) and $La_{1.8}Sr_{0.2}MMnO_6$

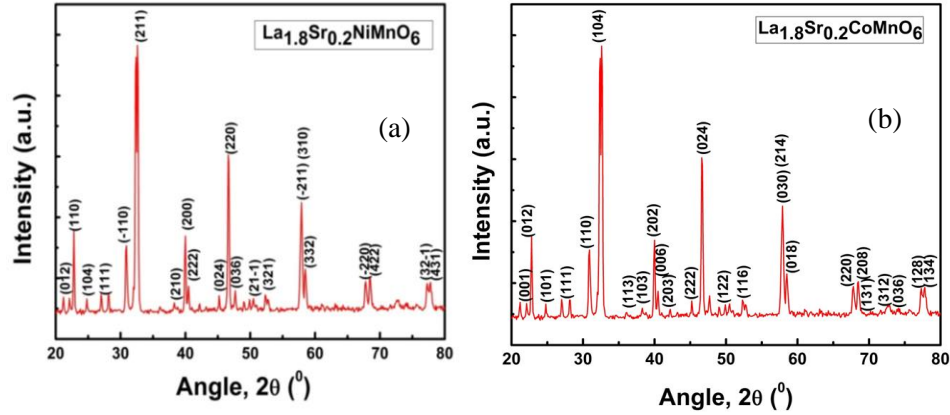

Figure 1. X-ray diffraction patterns at room temperature for $La_{1.8}Sr_{0.2}MMnO_6$ (M = Ni, Co). The peaks for all the samples can be well indexed by a simple phase with rhombohedral structure.

(M = Ni, Co) manganites are almost same, but lattice parameter 'c' is changed due to the 10% Sr substitution in La sites of $La_{1.8}Sr_{0.2}MMnO_6$ (M = Ni, Co) manganites.

Table 1: Lattice parameters of parent $La_2MMnO_6$ (M = Ni, Co) and $La_{1.8}Sr_{0.2}MMnO_6$ (M = Ni, Co) manganites.

| Parameters | $La_2NiMnO_6$ manganites (Å) [27, 28] | $La_{1.8}Sr_{0.2}NiMnO_6$ manganites (Å) [25, 29] | $La_2CoMnO_6$ manganites (Å) [26, 28, 30-31] | $La_{1.8}Sr_{0.2}CoMnO_6$ manganites (Å) [26] |
|---|---|---|---|---|
| A | 5.514 | 5.515 | 5.525 | 5.490 |
| B | 5.460 | 5.515 | 5.487 | 5.490 |
| C | 7.750 | 13.620 | 7.778 | 13.254 |

The Figure 2(a) shows the FESEM micrographs of the surface of bulk polycrystalline $La_{1.8}Sr_{0.2}NiMnO_6$ manganites sintered at 1300 °C. It is observed that the surface is non-homogeneous and the grains are agglomerated to some extent. Whereas, with the higher sintering temperature at 1400 °C, the sample became homogeneous and the grains were clearly visible (Figure 2(b)). The FESEM images of $La_{1.8}Sr_{0.2}CoMnO_6$ manganites sintered at 1300 °C (Figure 2(c)) reveal that particles are in good shape and grains are clear. However, for the same sample sintered at 1400 °C (Figure (d)), it is found that the homogeneity of the grains is destroyed. So, the optimum sintering temperatures are considered as 1400 °C and 1300 °C for $La_{1.8}Sr_{0.2}NiMnO_6$ and $La_{1.8}Sr_{0.2}CoMnO_6$ manganites, respectively. Figure 2(b) shows the histogram for average grain size (D) of $La_{1.8}Sr_{0.2}NiMnO_6$ manganite which is calculated by using ImageJ software from the FESEM micrographs and the D values are estimated to around 0.2~0.7 μm from the histogram. Notably, in the parent polycrystalline sample LNMO prepared by the solid state reaction method and sintered at 1400 °C, the average grain size estimated by FESEM image was 2 μm. [27] The histogram for grain size distribution of $La_{1.8}Sr_{0.2}CoMnO_6$ manganites which is worked out from the FESEM image (Figure 2(c)) and the average grain size is estimated at around 0.1~2.8 μm from the histogram. Noticeably, in the original polycrystalline sample

LCMO (sintered at 1300 °C) prepared by the same method, the average grain size was estimated to around 3-5 μm from FESEM image. [26]

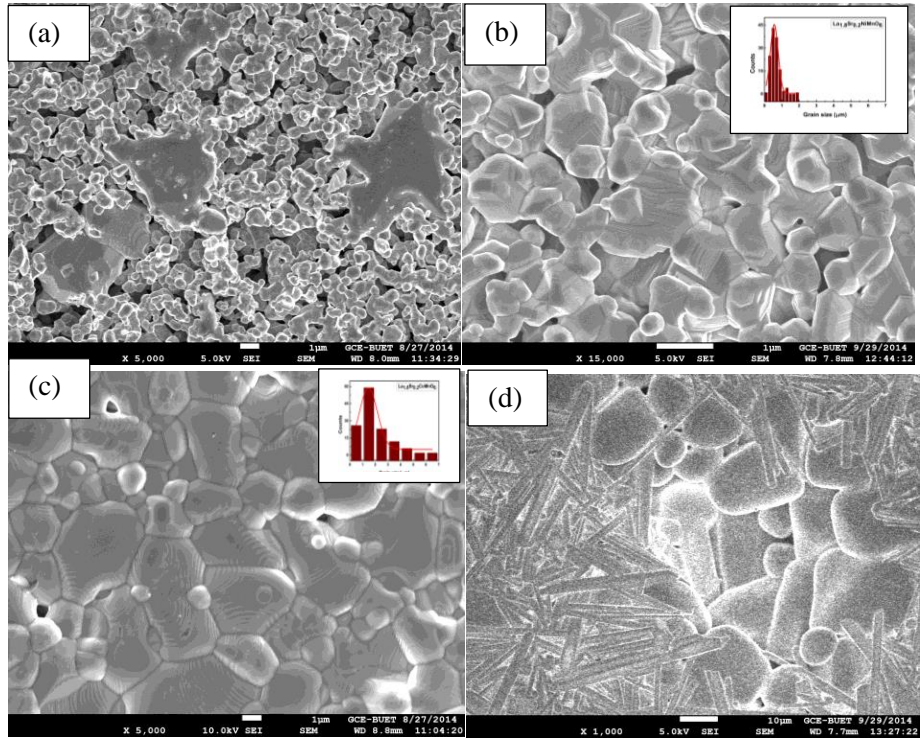

Figure 2. FESEM micrograph of $La_{1.8}Sr_{0.2}MMnO_6$ (M = Ni, Co) sintered at 1300 °C (a); (c) and 1400 °C (b); (d). The insets are the corresponding particle size histograms (b) and (c).

From the FESEM (Figure 2) micrographs, it is seen that the average grain size of parent $La_2MMnO_6$ (M = Ni, Co) manganites is greater than that of 10% Sr substitution in La sites of $La_{1.8}Sr_{0.2}MMnO_6$ (M = Ni, CO) manganites. The reduced grain size might be associated with different size of ionic radii caused the strain or stress field that hinders the grain growth process. It is clearly seen that the desired elements La, Sr, Ni, Mn and O of $La_{1.8}Sr_{0.2}NiMnO_6$ manganites (Figure 3(a)) and the desired elements La, Sr, Co, Mn and O of $La_{1.8}Sr_{0.2}CoMnO_6$ manganites (Figure 3(b)) are present.

Figures 4(a, b) show the XPS spectra of all elements of bulk $La_{1.8}Sr_{0.2}MMnO_6$ (M = Ni, Co) manganites. The photoemission data for all the samples was collected around the Mn 2p doublet, the La 4d and 3d doublet, the O 1s, Sr 3d, Co 2p, Ni 2p and C 1s lines. The XPS technique is a promising tool and is utilized to confirm the oxidation states of Mn, therefore, Mn states were explored elaborately and presented in Figures 4(c, d). Using high-resolution XPS, the two Mn 2p peaks at 642.54 and 654.26 eV, with an energy gap of 11.72 eV, can be assigned to Mn $2p_{3/2}$ and Mn $2p_{1/2}$, respectively as shown in Figure 4(c) for $La_{1.8}Sr_{0.2}NiMnO_6$ manganites. Similarly, the two Mn 2p peaks at 641.32 and 653.05 eV, with an energy gap of 11.73 eV, can be assigned to Mn $2p_{3/2}$ and Mn $2p_{1/2}$, respectively (Figure 4(d)) of $La_{1.8}Sr_{0.2}CoMnO_6$ manganites. However, it

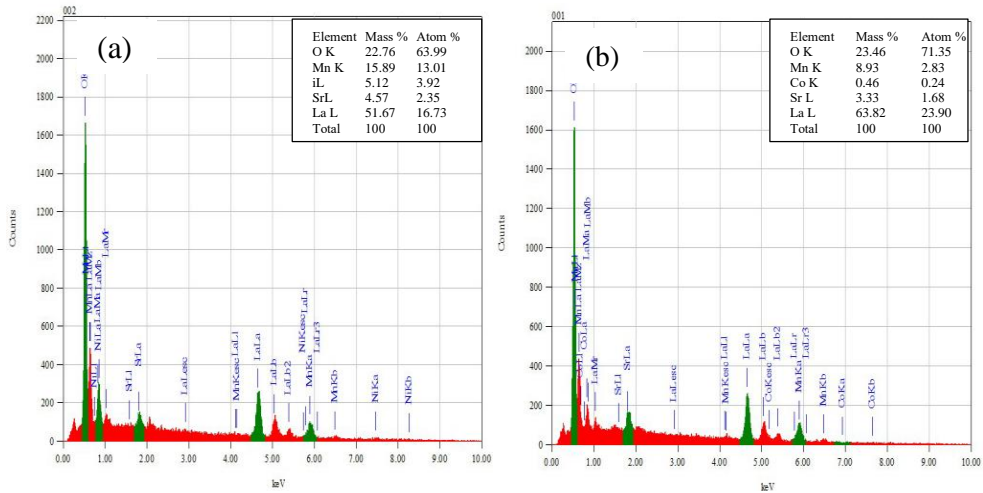

Figure 3. EDX spectrum of $La_{1.8}Sr_{0.2}MMnO_6$ (M = Ni, Co). The insets are the corresponding percentage of mass and atom of the sample.

is difficult to form a definite conclusion about the oxidation state of manganese from Mn 2p spectra, as all manganese oxidation states show the same pattern. Panda et al. [32], reported that the two Mn 2p peaks at 642.40 and 654.13 eV, with an energy gap of 11.73 eV, can be assigned to Mn $2p_{3/2}$ and Mn $2p_{1/2}$, respectively and Kang et al. [33], have reported that all the spectra display the spin-orbit split $2p_{3/2}$ and $2p_{1/2}$ peaks located around 642 and 654 eV, respectively which matches well with our reported present results. The Figure 4(e) illustrates the O 1s spectra of bulk $La_{1.8}Sr_{0.2}NiMnO_6$ manganites, which can be fitted to a symmetric Gaussian curve with peak positions 530.28 eV. On the contrary, in Figure 4(f), the O 1s spectra of bulk $La_{1.8}Sr_{0.2}CoMnO_6$ manganites show a slightly asymmetric peak very close to 528.10 eV with additional peak. The asymmetric curves of the bulk sample can be Gaussian fitted by two symmetrical peaks at 528.10 and 531.45 eV, respectively. The lower binding energy peak at 528.10 eV corresponds to the O 1s core spectrum, while higher binding energy peak is attributed to the oxygen vacancies, i.e., to the oxygen related defects [34-38] in the sample. In earlier investigations, a two-peak structure is observed with a component attributed to lattice oxygen at lower binding energies, and a component attributed to surface-or defect related oxygen at higher binding energies. [39-42] This interpretation is more consistent with the observations of the present work.

In the case of bulk $La_{1.8}Sr_{0.2}NiMnO_6$ manganites, temperature dependent zero field cooled and field cooled magnetization curves exhibit a sharp ferromagnetic (FM) to paramagnetic (PM) transition at 160±2 K under the application of a magnetic field of 500 Oe (Figures. 5(a, b)). But for the parent compound LNMO prepared by the same method, it is reported that temperature

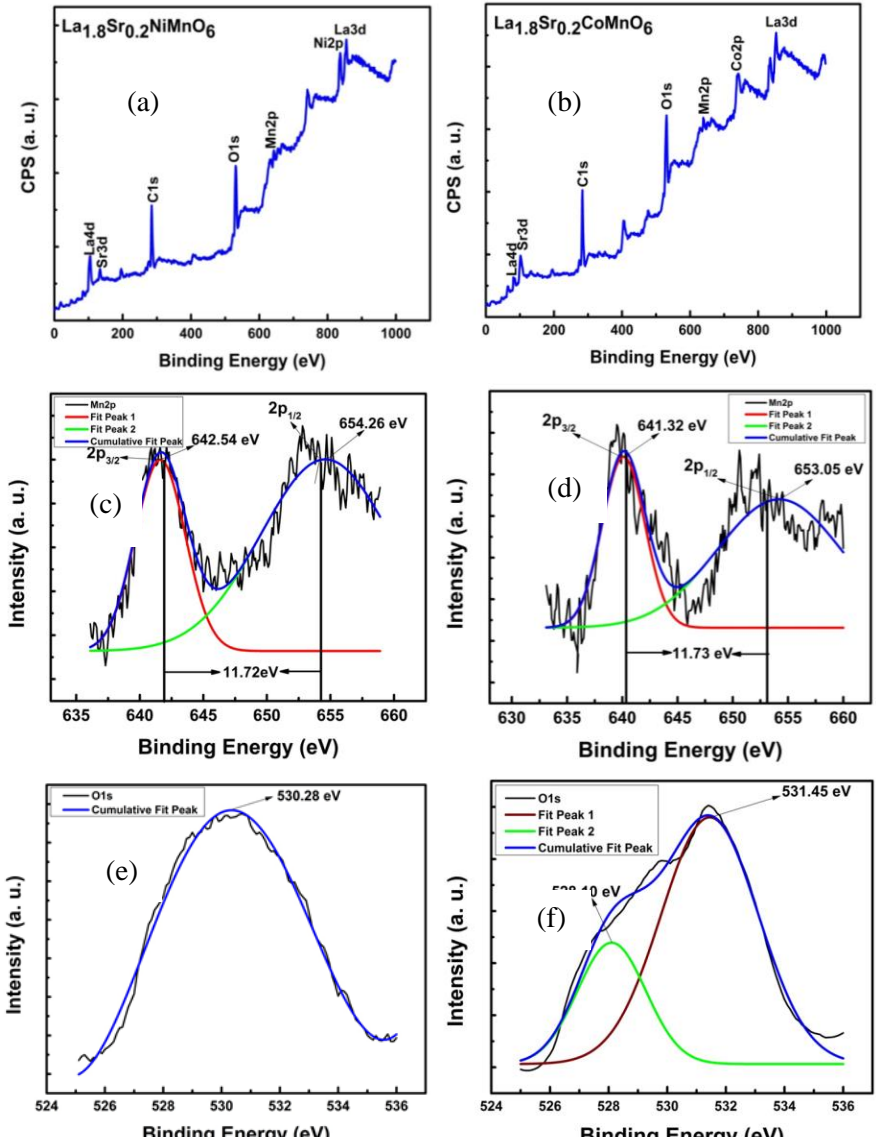

Figure 4. XPS spectra of all elements (a), (b); the Mn 2p (c), (d) and the O 1s (e), (f); of $La_{1.8}Sr_{0.2}MMnO_6$ (M = Ni, Co) manganites.

dependent ZFC and FC magnetization curves exhibit a sharp FM to PM transition at 280 K. [24, 27, 43-49] It has also been found that depending on the synthesis conditions, LNMO can exhibit one and more magnetic transitions. It is reported that LNMO shows two FM transitions occurring at $Tc_1$~280 K and $Tc_2$~150K. [7, 28, 43, 44, 50]

However, in the present investigation, we have observed a sharp transition at 220±2 K due to the substitution of Sr in LCMO manganites (Figures 5(c, d)). In the Figure 5(c), it is observed that there is a small dip in the plot around ~50 K that can be attributed to the presence of small amounts of hausmanite ($Mn_3O_4$), probably due to the presence of the higher electro positivity of Sr ions [51].

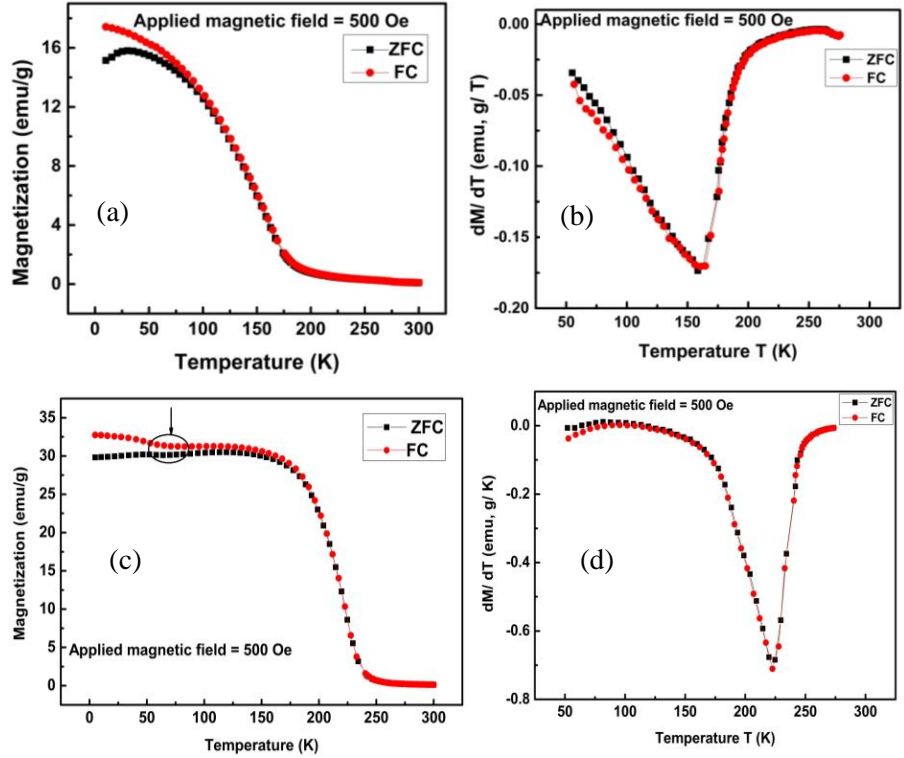

Figure 5. Temperature dependence of magnetization of $La_{1.8}Sr_{0.2}MMnO_6$ (M = Ni, Co) manganites Zero-Field-Cooled (ZFC) and Field-Cooled (FC) magnetization as a function of temperature (a) and (c). The first derivative of the temperature dependent magnetization (dM/dT) (b) and (d).

It is reported that the temperature dependence ZFC and FC magnetization curves exhibit a sharp FM to PM transition at 230 K for parent LCMO. [26] It has also been found that depending on the synthesis conditions LCMO can also exhibit one and more magnetic transitions. LCMO shows two FM transitions occurring at $T_{c1}$~220 K and $T_{c2}$~150 K. [52] Besides, other researchers reported that three distinct anomalies can be observed in both FC and ZFC curves of LCMO. The high-temperature anomaly is $T_{c1}$~210 K, the medium-temperature anomaly locates at $T_{c2}$~150 K, and the low-temperature anomaly is $T_{c3}$~80 K. [30] The multiple magnetic phase transitions have been reported earlier for both $La_2CoMnO_6$ ceramics and thin films. [7, 26, 28, 51-57] In this case, due to the substitution of Sr a sharp transition was observed at 220±2 K.

Notably, LNMO and LCMO can exhibit one or more magnetic transitions. In LNMO, an ordered sublattice with high spin $Ni^{2+}$ and $Mn^{4+}$ gives a FM transition at 280 K, while a disordered sublattice with low spin $Ni^{3+}$ and high spin $Mn^{3+}$ results in a FM transition below 150 K. [7, 28, 43, 44, 50] Similarly an ordered sublattice with high spin $Co^{2+}$ and $Mn^{4+}$ pairs in LCMO gives a FM transition at 220 K, while a disordered sublattice with low spin $Co^{3+}$ and high spin $Mn^{3+}$ results in a FM transition below 150 K. [52] The observed FM transition can be assigned to the superexchange interaction between $M^{2+}$-$O^{2-}$-$Mn^{4+}$ in $La_2MMnO_6$ (M = Ni, Co) manganites. However, the substitutions of La by Sr in LNMO and LCMO significantly affect the Curie temperature ($T_c$). The value of $T_c$ for LSNMO and LSCMO is determined by the FM coupling intensity between Ni and Mn; and Co and Mn ions which might be associated with the influence of modified structural parameters including the bond length and bond angle between $M^{2+}$-$O^{2-}$-$Mn^{4+}$.

It is evident that temperature dependent ZFC and FC magnetization curves of LSNMO exhibit a sharp FM to PM transition at 160±2 K which is around 120 K less than that of parent LNMO. On the other hand, temperature dependence ZFC and FC cooled magnetization curves of LSCMO exhibit a sharp FM to PM transition at 220±2 K which is around 8 K less than that of parent LCMO. The transition temperatures suggest that the effect of Sr on the transition temperature of LNMO is more significant than that in LCMO. Notably, microstructures obtained from FESEM images show that substitution of Sr in the $La_{1.8}Sr_{0.2}MMnO_6$ (M = Ni, Co) has demoted the grain growth process during sintering. We observed a correlation between the microstructural characteristics of LNMO and LCMO manganites and their magnetic transition temperatures. The temperature dependent magnetization measurements (Figure 5) demonstrated that the $T_c$ value is lower for Sr substituted LSNMO manganite for which the average grain size is smaller compared to that of LSCMO.

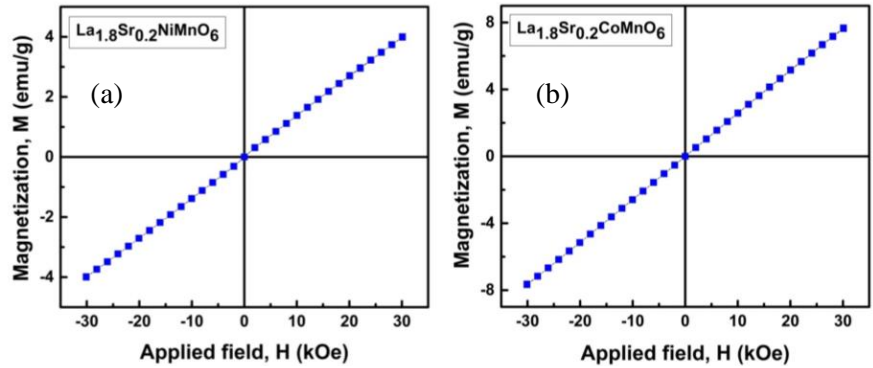

Figure 6. The room temperature M-H curves of $La_{1.8}Sr_{0.2}MMnO_6$ (M = Ni, Co) bulk materials.

The magnetization versus magnetic field (M-H) curves of $La_{1.8}Sr_{0.2}MMnO_6$ (M = Ni, Co) bulk materials were carried out at 300 K as represented in Figures 6(a, b). The unsaturated linear curves without any detectable hysteresis, clearly indicate the paramagnetic nature of the corresponding bulk sample.

**Conclusions**

The Sr substituted $La_{1.8}Sr_{0.2}MMnO_6$ (M = Ni, Co) manganites have been successfully synthesized and characterized. The average grain size of $La_{1.8}Sr_{0.2}NiMnO_6$ and $La_{1.8}Sr_{0.2}CoMnO_6$ have been found in the range 0.2~0.7 μm and 0.1~2.8 μm, respectively. The X-ray diffraction patterns revealed that Sr substituted $La_{1.8}Sr_{0.2}MMnO_6$ (M = Ni and Co) manganites has a rhombohedral structure. The chemical binding energies of the constituent elements of Sr substituted LSNMO and LSCMO manganites were confirmed from XPS analysis. For Sr substituted LSNMO, the FM to PM transition is found at around 160±2 K whilst for LSCMO, the transition is observed at around 220±2 K. The value of the transition temperature is lower for Sr substituted LSNMO manganite for which the average grain size is smaller compared to that of LSCMO. We may conclude that the addition of Sr in $La_{1.8}Sr_{0.2}MMnO_6$ manganites produced considerable modifications in the morphological, structural and magnetic properties of these manganites which might be useful in technological applications.

**Acknowledgments**

This work was supported by The Ministry of Science and Technology, Government of Bangladesh under research Grant No.: 39.009.002.01.00.053.2014-2015/PHY'S-273/ (26.01.2015). The magnetization measurements using SQUID were conducted in the Institute for Molecular Science (IMS), supported by Nanotechnology Platform Program (Molecule and Material Synthesis) of the Ministry of Education, Culture, Sports, Science and Technology (MEXT), Japan.